\documentclass[11pt, a4paper]{article}

\usepackage{jheppub}

\newcommand{\ChS}{\text{CS}}
\newcommand{\gWZW}{\text{gWZW}}

\newcommand{\TN}{\mathrm{TN}}

\newcommand{\gf}{\mathfrak{g}}
\newcommand{\gfh}{\widehat{\gf}}

\newcommand{\del}{\partial}
\newcommand{\delb}{{\bar\partial}}

\newcommand{\vev}[1]{\langle #1 \rangle}

\newcommand{\rank}{\mathop{\mathrm{rank}}\nolimits}
\renewcommand{\Im}{\mathop{\mathrm{Im}}\nolimits}
\renewcommand{\Re}{\mathop{\mathrm{Re}}\nolimits}

\newcommand{\Tr}{\mathop{\mathrm{Tr}}\nolimits}

\newcommand{\SU}{\mathrm{SU}}
\newcommand{\SO}{\mathrm{SO}}
\newcommand{\Spin}{\mathrm{Spin}}

\newcommand{\slf}{\mathfrak{sl}}
\newcommand{\suf}{\mathfrak{su}}

\newcommand{\U}{\mathrm{U}}

\newcommand{\iso}{\cong}

\newcommand{\Z}{\mathbb{Z}}

\newcommand{\R}{\mathbb{R}}
\newcommand{\C}{\mathbb{C}}

\let\nc\newcommand
\let\renc\renewcommand
\nc{\wbar}{\overline}
\let\td\tilde
\let\wtd\widetilde
\let\wht\widehat
\let\mcl\mathcal

\nc{\ab}{{\bar{a}}} \nc{\at}{\tilde{a}} \nc{\ah}{\hat{a}}
\nc{\bb}{{\bar{b}}} \nc{\bt}{\tilde{b}} \nc{\bh}{\hat{b}}
\nc{\cb}{{\bar{c}}} \nc{\ct}{\tilde{c}} 
\nc{\db}{{\bar{d}}} \nc{\dt}{\tilde{d}} \renc{\dh}{\hat{d}}
\nc{\eb}{{\bar{e}}} \nc{\et}{\tilde{e}} \nc{\eh}{\hat{e}}
\nc{\fb}{{\bar{f}}} \nc{\ft}{\tilde{f}} \nc{\fh}{\hat{f}}
\nc{\gb}{{\bar{g}}} \nc{\gt}{\tilde{g}} \nc{\gh}{\hat{g}}
\nc{\hb}{{\bar{h}}} \nc{\hh}{\hat{h}} 
\nc{\ib}{{\bar{\imath}}} \nc{\ih}{\hat{\imath}} 
\nc{\jb}{{\bar{\jmath}}} \nc{\jt}{\tilde{\jmath}} \nc{\jh}{\hat{\jmath}}
\nc{\kb}{{\bar{k}}} \nc{\kt}{\tilde{k}} \nc{\kh}{\hat{k}}
\nc{\lb}{{\bar{l}}} \nc{\lt}{\tilde{l}} \nc{\lh}{\hat{l}}
\nc{\mb}{{\bar{m}}} \nc{\mt}{\tilde{m}} \nc{\mh}{\hat{m}}
\nc{\nb}{{\bar{n}}} \nc{\nt}{\tilde{n}} \nc{\nh}{\hat{n}}
\nc{\ob}{{\bar{o}}} \nc{\ot}{\tilde{o}} \nc{\oh}{\hat{o}}
\nc{\pb}{{\bar{p}}} \nc{\pt}{\tilde{p}} \nc{\ph}{\hat{p}}
\nc{\qb}{{\bar{q}}} \nc{\qt}{\tilde{q}} \nc{\qh}{\hat{q}}
\nc{\rb}{{\bar{r}}} \nc{\rt}{\tilde{r}} \nc{\rh}{\hat{r}}
\renc{\sb}{{\bar{s}}} \nc{\st}{\tilde{s}} \nc{\sh}{\hat{s}}
\nc{\tb}{{\bar{t}}} \renc{\th}{\hat{t}} 
\nc{\ub}{{\bar{u}}} \nc{\ut}{\tilde{u}} \nc{\uh}{\hat{u}}
\nc{\vb}{{\bar{v}}} \nc{\vt}{\tilde{v}} \nc{\vh}{\hat{v}}
\nc{\wb}{{\bar{w}}} \nc{\wt}{\tilde{w}} \nc{\wh}{\hat{w}}
\nc{\xb}{{\bar{x}}} \nc{\xt}{\tilde{x}} \nc{\xh}{\hat{x}}
\nc{\yb}{{\bar{y}}} \nc{\yt}{\tilde{y}} \nc{\yh}{\hat{y}}
\nc{\zb}{{\bar{z}}} \nc{\zt}{\tilde{z}} \nc{\zh}{\hat{z}}

\nc{\Ab}{\wbar{A}} \nc{\At}{\wtd{A}} \nc{\Ah}{\wht{A}}
\nc{\Bb}{\wbar{B}} \nc{\Bt}{\wtd{B}} \nc{\Bh}{\wht{B}}
\nc{\Cb}{\wbar{C}} \nc{\Ct}{\wtd{C}} \nc{\Ch}{\wht{C}}
\nc{\Db}{\wbar{D}} \nc{\Dt}{\wtd{D}} \nc{\Dh}{\wht{D}}
\nc{\Eb}{\wbar{E}} \nc{\Et}{\wtd{E}} \nc{\Eh}{\wht{E}}
\nc{\Fb}{\wbar{F}} \nc{\Ft}{\wtd{F}} \nc{\Fh}{\wht{F}}
\nc{\Gb}{\wbar{G}} \nc{\Gt}{\wtd{G}} \nc{\Gh}{\wht{G}}
\nc{\Hb}{\wbar{H}} \nc{\Ht}{\wtd{H}} \nc{\Hh}{\wht{H}}
\nc{\Ib}{\wbar{I}} \nc{\It}{\wtd{I}} \nc{\Ih}{\wht{I}}
\nc{\Jb}{\wbar{J}} \nc{\Jt}{\wtd{J}} \nc{\Jh}{\wht{J}}
\nc{\Kb}{\wbar{K}} \nc{\Kt}{\wtd{K}} \nc{\Kh}{\wht{K}}
\nc{\Lb}{\wbar{L}} \nc{\Lt}{\wtd{L}} \nc{\Lh}{\wht{L}}
\nc{\Mb}{\wbar{M}} \nc{\Mt}{\wtd{M}} \nc{\Mh}{\wht{M}}
\nc{\Nb}{\wbar{N}} \nc{\Nt}{\wtd{N}} \nc{\Nh}{\wht{N}}
\nc{\Ob}{\wbar{O}} \nc{\Ot}{\wtd{O}} \nc{\Oh}{\wht{O}}
\nc{\Pb}{\wbar{P}} \nc{\Pt}{\wtd{P}} \nc{\Ph}{\wht{P}}
\nc{\Qb}{\wbar{Q}} \nc{\Qt}{\wtd{Q}} \nc{\Qh}{\wht{Q}}
\nc{\Rb}{\wbar{R}} \nc{\Rt}{\wtd{R}} \nc{\Rh}{\wht{R}}
\nc{\Sb}{\wbar{S}} \nc{\St}{\wtd{S}} \nc{\Sh}{\wht{S}}
\nc{\Tb}{\wbar{T}} \nc{\Tt}{\wtd{T}} \nc{\Th}{\wht{T}}
\nc{\Ub}{\wbar{U}} \nc{\Ut}{\wtd{U}} \nc{\Uh}{\wht{U}}
\nc{\Vb}{\wbar{V}} \nc{\Vt}{\wtd{V}} \nc{\Vh}{\wht{V}}
\nc{\Wb}{\wbar{W}} \nc{\Wt}{\wtd{W}} \nc{\Wh}{\wht{W}}
\nc{\Xb}{\wbar{X}} \nc{\Xt}{\wtd{X}} \nc{\Xh}{\wht{X}}
\nc{\Yb}{\wbar{Y}} \nc{\Yt}{\wtd{Y}} \nc{\Yh}{\wht{Y}}
\nc{\Zb}{\wbar{Z}} \nc{\Zt}{\wtd{Z}} \nc{\Zh}{\wht{Z}}

\nc{\CA}{\mcl{A}} \nc{\CAb}{\wbar{\CA}} \nc{\CAt}{\wtd{\CA}} \nc{\CAh}{\wht{\CA}}
\nc{\CB}{\mcl{B}} \nc{\CBb}{\wbar{\CB}} \nc{\CBt}{\wtd{\CB}} \nc{\CBh}{\wht{\CB}}
\nc{\CC}{\mcl{C}} \nc{\CCb}{\wbar{\CC}} \nc{\CCt}{\wtd{\CC}} \nc{\CCh}{\wht{\CC}}
\nc{\CD}{\mcl{D}} \nc{\CDb}{\wbar{\CD}} \nc{\CDt}{\wtd{\CD}} \nc{\CDh}{\wht{\CD}}
\nc{\CE}{\mcl{E}} \nc{\CEb}{\wbar{\CE}} \nc{\CEt}{\wtd{\CE}} \nc{\CEh}{\wht{\CE}}
\nc{\CF}{\mcl{F}} \nc{\CFb}{\wbar{\CF}} \nc{\CFt}{\wtd{\CF}} \nc{\CFh}{\wht{\CF}}
\nc{\CG}{\mcl{G}} \nc{\CGb}{\wbar{\CG}} \nc{\CGt}{\wtd{\CG}} \nc{\CGh}{\wht{\CG}}
\nc{\CH}{\mcl{H}} \nc{\CHb}{\wbar{\CH}} \nc{\CHt}{\wtd{\CH}} \nc{\CHh}{\wht{\CH}}
\nc{\CI}{\mcl{I}} \nc{\CIb}{\wbar{\CI}} \nc{\CIt}{\wtd{\CI}} \nc{\CIh}{\wht{\CI}}
\nc{\CJ}{\mcl{J}} \nc{\CJb}{\wbar{\CJ}} \nc{\CJt}{\wtd{\CJ}} \nc{\CJh}{\wht{\CJ}}
\nc{\CK}{\mcl{K}} \nc{\CKb}{\wbar{\CK}} \nc{\CKt}{\wtd{\CK}} \nc{\CKh}{\wht{\CK}}
\nc{\CL}{\mcl{L}} \nc{\CLb}{\wbar{\CL}} \nc{\CLt}{\wtd{\CL}} \nc{\CLh}{\wht{\CL}}
\nc{\CM}{\mcl{M}} \nc{\CMb}{\wbar{\CM}} \nc{\CMt}{\wtd{\CM}} \nc{\CMh}{\wht{\CM}}
\nc{\CN}{\mcl{N}} \nc{\CNb}{\wbar{\CN}} \nc{\CNt}{\wtd{\CN}} \nc{\CNh}{\wht{\CN}}
\nc{\CO}{\mcl{O}} \nc{\COb}{\wbar{\CO}} \nc{\COt}{\wtd{\CO}} \nc{\COh}{\wht{\CO}}
\nc{\CQ}{\mcl{Q}} \nc{\CQb}{\wbar{\CQ}} \nc{\CQt}{\wtd{\CQ}} \nc{\CQh}{\wht{\CQ}}
\nc{\CR}{\mcl{R}} \nc{\CRb}{\wbar{\CR}} \nc{\CRt}{\wtd{\CR}} \nc{\CRh}{\wht{\CR}}
\nc{\CS}{\mcl{S}} \nc{\CSb}{\wbar{\CS}} \nc{\CSt}{\wtd{\CS}} \nc{\CSh}{\wht{\CS}}
\nc{\CT}{\mcl{T}} \nc{\CTb}{\wbar{\CT}} \nc{\CTt}{\wtd{\CT}} \nc{\CTh}{\wht{\CT}}
\nc{\CU}{\mcl{U}} \nc{\CUb}{\wbar{\CU}} \nc{\CUt}{\wtd{\CU}} \nc{\CUh}{\wht{\CU}}
\nc{\CV}{\mcl{V}} \nc{\CVb}{\wbar{\CV}} \nc{\CVt}{\wtd{\CV}} \nc{\CVh}{\wht{\CV}}
\nc{\CW}{\mcl{W}} \nc{\CWb}{\wbar{\CW}} \nc{\CWt}{\wtd{\CW}} \nc{\CWh}{\wht{\CW}}
\nc{\CX}{\mcl{X}} \nc{\CXb}{\wbar{\CX}} \nc{\CXt}{\wtd{\CX}} \nc{\CXh}{\wht{\CX}}
\nc{\CY}{\mcl{Y}} \nc{\CYb}{\wbar{\CY}} \nc{\CYt}{\wtd{\CY}} \nc{\CYh}{\wht{\CY}}
\nc{\CZ}{\mcl{Z}} \nc{\CZb}{\wbar{\CZ}} \nc{\CZt}{\wtd{\CZ}} \nc{\CZh}{\wht{\CZ}}

\let\eps\epsilon
\let\ups\upsilon
\let\veps\varepsilon
\let\vtht\vartheta
\let\vsgm\varsigma
\let\vphi\varphi
\let\vrho\varrho

\nc{\alphab}{\bar{\alpha}} \nc{\alphat}{\td{\alpha}} \nc{\alphah}{\hat{\alpha}}
\nc{\betab}{\bar{\beta}}   \nc{\betat}{\td{\beta}}   \nc{\betah}{\hat{\beta}} 
\nc{\gammab}{\bar{\gamma}} \nc{\gammat}{\td{\gamma}} \nc{\gammah}{\hat{\gamma}} 
\nc{\deltab}{\bar{\delta}} \nc{\deltat}{\td{\delta}} \nc{\deltah}{\hat{\delta}} 
\nc{\epsilonb}{\bar{\eps}} \nc{\epsilont}{\td{\eps}} \nc{\epsilonh}{\hat{\eps}} 
\nc{\vepsb}{\bar{\veps}}   \nc{\vepst}{\td{\veps}}   \nc{\vepsh}{\hat{\veps}} 
\nc{\zetab}{\bar{\zeta}}   \nc{\zetat}{\td{\zeta}}   \nc{\zetah}{\hat{\zeta}} 
\nc{\etab}{\bar{\eta}}     \nc{\etat}{\td{\eta}}     \nc{\etah}{\hat{\eta}} 
\nc{\thetab}{\bar{\theta}} \nc{\thetat}{\td{\theta}} \nc{\thetah}{\hat{\theta}} 
\nc{\vthetab}{\bar{\vtht}} \nc{\vthetat}{\td{\vtht}} \nc{\vthetah}{\hat{\vtht}} 
\nc{\lambdab}{\bar{\lambda}} \nc{\lambdat}{\td{\lambda}} \nc{\lambdah}{\hat{\lambda}} 
\nc{\iotab}{\bar{\iota}}   \nc{\iotat}{\td{\iota}}   \nc{\iotah}{\hat{\iota}} 
\nc{\kappab}{\bar{\kappa}} \nc{\kappat}{\td{\kappa}} \nc{\kappah}{\hat{\kappa}} 
\nc{\lmdb}{\bar{\lmd}}     \nc{\lmdt}{\td{\lmd}}     \nc{\lmdh}{\hat{\lmd}} 
\nc{\mub}{\bar{\mu}}       \nc{\mut}{\td{\mu}}       \nc{\muh}{\hat{\mu}} 
\nc{\nub}{\bar{\nu}}       \nc{\nut}{\td{\nu}}       \nc{\nuh}{\hat{\nu}} 
\nc{\xib}{\bar{\xi}}       \nc{\xit}{\td{\xi}}       \nc{\xih}{\hat{\xi}} 
\nc{\pib}{\bar{\pi}}       \nc{\pit}{\td{\pi}}       \nc{\pih}{\hat{\pi}} 
\nc{\vpib}{\bar{\vpi}}     \nc{\vpit}{\td{\vpi}}     \nc{\vpih}{\hat{\vpi}} 
\nc{\rhob}{\bar{\rho}}     \nc{\rhot}{\td{\rho}}     \nc{\rhoh}{\hat{\rho}} 
\nc{\vrhob}{\bar{\vrho}}   \nc{\vrhot}{\td{\vrho}}   \nc{\vrhoh}{\hat{\vrho}} 
\nc{\sigmab}{\bar{\sigma}} \nc{\sigmat}{\td{\sigma}} \nc{\sigmah}{\hat{\sigma}} 
\nc{\vsigmab}{\bar{\vsgm}} \nc{\vsigmat}{\td{\vsgm}} \nc{\vsigmah}{\hat{\vsgm}} 
\nc{\taub}{\bar{\tau}}     \nc{\taut}{\td{\tau}}     \nc{\tauh}{\hat{\tau}} 
\nc{\upsilonb}{\bar{\ups}} \nc{\upsilont}{\td{\ups}} \nc{\upsilonh}{\hat{\ups}} 
\nc{\phib}{\bar{\phi}}     \nc{\phit}{\td{\phi}}     \nc{\phih}{\hat{\phi}} 
\nc{\varphib}{\bar{\vphi}}   \nc{\varphit}{\td{\vphi}}   \nc{\varphih}{\hat{\vphi}} 
\nc{\chib}{\bar{\chi}}     \nc{\chit}{\td{\chi}}     \nc{\chih}{\hat{\chi}} 
\nc{\psib}{\bar{\psi}}     \nc{\psit}{\td{\psi}}     \nc{\psih}{\hat{\psi}} 
\nc{\omegab}{\bar{\omega}} \nc{\omegat}{\td{\omega}} \nc{\omegah}{\hat{\omega}} 

\nc{\Gammab}{\wbar{\Gamma}}     \nc{\Gammat}{\wtd{\Gamma}}     \nc{\Gammah}{\wht{\Gamma}}
\nc{\Deltab}{\wbar{\Delta}}     \nc{\Deltat}{\wtd{\Delta}}     \nc{\Deltah}{\wht{\Delta}}
\nc{\Thetab}{\wbar{\Theta}}     \nc{\Thetat}{\wtd{\Theta}}     \nc{\Thetah}{\wht{\Theta}}
\nc{\Lambdab}{\wbar{\Lambda}}   \nc{\Lambdat}{\wtd{\Lambda}}   \nc{\Lambdah}{\wht{\Lambda}}
\nc{\Xib}{\wbar{\Xi}}           \nc{\Xit}{\wtd{\Xi}}           \nc{\Xih}{\wht{\Xi}}
\nc{\Pib}{\wbar{\Pi}}           \nc{\Pit}{\wtd{\Pi}}           \nc{\Pih}{\wht{\Pi}}
\nc{\Sigmab}{\wbar{\Sigma}}     \nc{\Sigmat}{\wtd{\Sigma}}     \nc{\Sigmah}{\wht{\Sigma}}
\nc{\Upsilonb}{\wbar{\Upsilon}} \nc{\Upsilont}{\wtd{\Upsilon}} \nc{\Upsilonh}{\wht{\Upsilon}}
\nc{\Phib}{\wbar{\Phi}}         \nc{\Phit}{\wtd{\Phi}}         \nc{\Phih}{\wht{\Phi}}
\nc{\Psib}{\wbar{\Psi}}         \nc{\Psit}{\wtd{\Psi}}         \nc{\Psih}{\wht{\Psi}}
\nc{\Omegab}{\wbar{\Omega}}     \nc{\Omegat}{\wtd{\Omega}}     \nc{\Omegah}{\wht{\Omega}}

\makeatletter
\def\wbar{\accentset{{\cc@style\underline{\mskip12mu}}}}
\def\wbarl{\accentset{{\cc@style\mskip-2mu\underline{\mskip12mu}}}}
\let\wb@r\wbar
\let\wb@rl\wbarl
\renewcommand{\wbar}[1]{\wb@r{#1}}
\renewcommand{\wbarl}[1]{\wb@rl{#1}}
\renewcommand{\Tb}{\wbarl{T}}
\renewcommand{\Qb}{\wbarl{Q}}
\makeatother

\renewcommand{\psit}{\tilde\psi}
\renewcommand{\psib}{\bar\psi}

\title{Compactification on the \texorpdfstring{$\mathbf\Omega$}{Omega}-background and the AGT correspondence}
\author{Junya Yagi}
\affiliation{Department of Mathematics, University of Hamburg, \\
Bundesstra\ss e 55, 20146 Hamburg, Germany}
\emailAdd{junya.yagi@math.uni-hamburg.de}

\abstract{The six-dimensional $(2,0)$ theory formulated in the
  $\Omega$-background gives rise to two-dimensional effective degrees
  of freedom.  By compactifying the theory on the circle fibers of two
  cigar-like manifolds, we find that a natural candidate for the
  effective theory is a chiral gauged WZW model.  The symmetry algebra
  of the model contains the W-algebra that appears on the
  two-dimensional side of the AGT correspondence.  We show that the
  expectation values of its currents determine the Seiberg-Witten
  curve of the four-dimensional side.}


\arxivnumber{1205.6820}

\begin{document}
\maketitle
\flushbottom

\section{Introduction}

The AGT correspondence \cite{Alday:2009aq} relates $\CN = 2$
supersymmetric gauge theories in four dimensions and conformal field
theories with W-algebra symmetry in two dimensions
\cite{Mironov:2009by, Wyllard:2009hg, Taki:2009zd, Alday:2010vg,
  Kozcaz:2010yp, Braverman:2010ef, Wyllard:2010rp}.  Various objects
from the two sides are identified through this correspondence, such as
the Nekrasov partition functions and conformal blocks
\cite{Alday:2009aq}, the generators of the chiral rings and the
currents of the W-algebras \cite{Bonelli:2009zp}, and the
Seiberg-Witten curves and the expectation values of the W-currents
\cite{Alday:2009aq}.  Perhaps more fundamentally, the correspondence
is also manifested in the existence of a W-algebra action on the
equivariant cohomology of the instanton moduli space \cite{MO,
  Schiffmann:2012}.

It has been argued that the connection between these seemingly distant
theories originates from six dimensions.  The starting point is the
$(2,0)$ theory on $M \times C$, with $M$ a four-manifold and $C$ a
punctured Riemann surface, and codimension-two defect operators placed
at the punctures of $C$.  For a general choice of the product metric
supersymmetry is completely broken, but one can twist the theory to
save one of the sixteen supercharges; call it $Q$.  When this is done,
the theory is expected to become topological along $M$ and holomorphic
along $C$ in the $Q$-invariant sector.  If one compactifies this
twisted $(2,0)$ theory on $C$, one gets an $\CN = 2$ gauge theory on
$M$ \cite{Witten:1997sc, Gaiotto:2009we, Gaiotto:2009hg} with the
familiar Donaldson-Witten twist \cite{Witten:1988ze}.  If one
compactifies the theory instead on $M$, then one ends up with a
twisted $\CN = (0,2)$ supersymmetric theory on $C$, which in the
present case will be a chiral conformal field theory.  In the twisted
theory, physical quantities are protected under rescaling of the
metric of $M$ or $C$.  Then, comparing the effective descriptions of
protected quantities leads to a correspondence between the four- and
two-dimensional theories.

This argument ignores a crucial point, however: the AGT correspondence
does not deal with $\CN = 2$ gauge theories of the standard type with
$M$ compact, but rather involves the $\Omega$-deformation
\cite{Nekrasov:2002qd} of them with $M$ noncompact.  Thus, one must
really consider the situation where the above setup is subject to a
deformation that reduces to the $\Omega$-deformation upon
compactification on $C$.  In the works \cite{Alday:2009qq,
  Tachikawa:2011dz, Nishioka:2011jk} where the central charges of the
effective conformal field theories were computed from anomalies of the
$(2,0)$ theory, the effect of the $\Omega$-deformation was
incorporated by replacing anomaly polynomials by their equivariant
counterparts.  The success of this procedure indicates that such a
deformation does exist.  In fact, an M-theory construction has been
proposed recently \cite{Hellerman:2012zf}.

The goal of this paper is to understand how the expected conformal
field theories arise at low energies in the case $M = \R^4$, assuming
that there is a formulation of the twisted $(2,0)$ theory in the
$\Omega$-background in the sense just described.

In the standard four-dimensional formulation, the $\Omega$-deformation
confines quantum effects near the origin of $\R^4$, within a region
whose characteristic scale is set by the deformation parameters.  So
by taking the parameters to be large compared to the energy scale of
interest, one can localize quantum effects to the origin.  One may
call this procedure ``compactification on the $\Omega$-background.''
If one applies the same procedure to the $(2,0)$ theory on $\R^4
\times C$, one should obtain an effective theory on $C$ describing
degrees of freedom living at the origin of $\R^4$.

To identify this effective theory, we perform a different
compactification.  Exploiting the quasi-topological nature of the
twisted theory, we bend $\R^4$ into the product $D_1 \times D_2$ of
two cigar-like manifolds, each consisting of a semi-infinite cylinder
capped with a hemisphere on one end, and then take the radii of the
cigars to be small.  A peculiar feature of this geometry is that on
the flat cylinder region the $\Omega$-deformation can be canceled by a
change of variables \cite{Nekrasov:2010ka}.  This property allows us
to represent the effect of the $\Omega$-deformation by the insertion
of $Q$-invariant operators, supported on codimension-two submanifolds
located at the tips of the cigars.  We describe how this works in
section~\ref{Omega}.

In the absence of the $\Omega$-deformation, compactification on the
circle fibers of the cigars gives $\CN = 4$ super Yang-Mills theory on
$L_1 \times L_2 \times C$, where $L_1$ and $L_2$ are respectively the
\linebreak axes of $D_1$ and $D_2$, each a half-line $[0,\infty)$.  This
four-manifold with a corner has two boundary components intersecting
orthogonally, $\{0\} \times L_2 \times C$ coming from the tip of $D_1$
and $L_1 \times \linebreak \{0\} \times C$ coming from the tip of $D_2$.  These
boundaries are endowed with half-BPS boundary conditions related by
$S$-duality.  We determine these boundary conditions in
section~\ref{4d}.

Turning on the $\Omega$-deformation translates in the compactified
theory to introducing $Q$-invariant boundary terms to the action.
This may sound strange.  After all, the $\Omega$-deformation is
expected to give rise to two-dimensional dynamics, not three.
Shouldn't it then produce something defined on a two-dimensional
submanifold?  Quite the contrary, these boundary couplings generate
exactly such dynamics, because the two boundaries themselves have a
common boundary which is two-dimensional.  The boundary couplings must
satisfy certain criteria derived from the quasi-topological invariance
in six dimensions.  In section \ref{WZW}, we will find that natural
candidates are a Chern-Simons term for the complexified gauge group
and its $S$-dual, which, formulated on a manifold with boundary,
induce boundary degrees of freedom described by a gauged WZW model
\cite{Witten:1988hf, Moore:1989yh, Elitzur:1989nr}.  The emergence of
dynamical boundary degrees of freedom from otherwise topological
Chern-Simons theory is a prototypical example of holography.

These degrees of freedom come from six dimensions since
compactification cannot cre\-ate new ones.  In turn, they should be the
degrees of freedom seen at low energies of the $(2,0)$ theory
formulated in the $\Omega$-background --- the intersection of the two
boundaries is nothing but the origin of $\R^4$!  Therefore, we
conclude that the $(2,0)$ theory ``compactified on the
$\Omega$-background'' is described by the gauged WZW model.  The
purely bosonic nature of the latter theory solves an apparent puzzle
that conformal field theories relevant for the AGT correspondence for
$M = \R^4$ are not supersymmetric as the naive compactification
argument may suggest.

Besides, the model has the right property: its symmetry algebra
contains a W-algebra, precisely the one that appears in the AGT
correspondence.  Furthermore, one can relate the W-algebra quite
naturally to the Seiberg-Witten curve of the effective $\CN = 2$
theory, and the relation agrees with the conjectured one.  These
facts, explained more fully in section \ref{AGT}, provide evidence for
the correctness of our argument.

Our conclusion reinforces the idea that the $Q$-cohomology of the
$(2,0)$ theory contains a W-algebra \cite{Yagi:2011vd}, albeit $Q$
here should probably be replaced by a different supercharge $\Qt$
appropriate for the $\Omega$-deformed situation.  Up in six
dimensions, the $\Qt$-cohomology of states is naturally a module over
the $\Qt$-cohomology of operators.  Once we go down to four
dimensions, the action of some subalgebra may still be present, but
then the way it is realized is probably not very obvious.  The AGT
correspondence seems to be an example of such an instance: the
$\Qt$-cohomology in six dimensions contains the W-algebra, and it acts
on the $\Qt$-cohomology of the $\Omega$-deformed $\CN = 2$ theory ---
or the equivariant cohomology of the instanton moduli space --- in a
nontrivial way.

\section{Localizing the \texorpdfstring{$\mathbf\Omega$}{Omega}-deformation}
\label{Omega}

To begin, let us explain how the $(2,0)$ theory on $M \times C$ is
twisted, and why the twisted theory is expected to be topological
along $M$ and holomorphic along $C$.

The general idea of twisting is to replace the holonomy group by a
combination of the holonomy and R-symmetry groups under which a
fraction of the supercharges transform as scalars, therefore are left
unbroken by the curvature.  For the $(2,0)$ theory on $M \times C$,
the holonomy is reduced to
\begin{equation}
  \Spin(4) \times \Spin(2) \iso \SU(2)_\ell \times \SU(2)_{r} \times \U(1)_C,
\end{equation}
and the R-symmetry is $\Spin(5)$.  We split the latter as
\begin{equation}
  \Spin(3) \times \Spin(2) \iso \SU(2)_R \times \U(1)_\CR.
\end{equation}
The supercharges transform as $\mathbf{4}_+ \otimes \mathbf{4}$ under
$\Spin(6) \times \Spin(5)$, where $\mathbf{4}_+$ is a positive
chirality spinor of $\Spin(6)$ and $\mathbf{4}$ is a spinor of
$\Spin(5)$.  Under the above subgroups of $\Spin(6)$ and $\Spin(5)$,
they decompose as
\begin{equation}
  \mathbf 4_+ \otimes \mathbf 4
  \to \bigl((\mathbf 2, \mathbf 1)_{1/2}
      \oplus  (\mathbf 1, \mathbf 2)_{-1/2}\bigr)
      \otimes (\mathbf 2_{{1/2}} \oplus \mathbf 2_{-1/2}).
\end{equation}
We replace $\SU(2)_r$ by the diagonal subgroup $\SU(2)_r'$ of
$\SU(2)_r \times \SU(2)_R$, and $\U(1)_C$ by the diagonal subgroup
$\U(1)_C'$ of $\U(1)_C \times \U(1)_\CR$.  Under $\SU(2)_\ell \times
\SU(2)_r' \times \U(1)_C'$, the supercharges transform as
\begin{equation}
  (\mathbf 2, \mathbf 2)_{1} \oplus (\mathbf 2, \mathbf 2)_0
  \oplus (\mathbf 1, \mathbf 1)_0 \oplus (\mathbf 1, \mathbf 3)_0
  \oplus (\mathbf 1, \mathbf 1)_{-1} \oplus (\mathbf 1, \mathbf 3)_{-1}.
\end{equation}
We see that the twisting produces one scalar supercharge, which we
call $Q$.  From the viewpoint of $M$, this is the Donaldson-Witten
twist \cite{Witten:1988ze} of $\CN = 2$ supersymmetry.  From the
viewpoint of $C$, it is the unique twist of $\CN = (0,2)$
supersymmetry.

In the case of twisted $\CN = 2$ gauge theories in four dimensions, a
similar supercharge obeys $Q^2 = 0$ up to a gauge transformation, and
upon restricting to $Q$- and gauge-invariant operators and states,
physical quantities depend only on the $Q$-cohomology classes of
operators and states involved.  Meanwhile, the energy-momentum tensor
is $Q$-exact.  It follows that the twisted theories are topological,
in the sense that physical quantities are invariant under deformations
of the metric.  Likewise, for twisted $\CN = (0,2)$ theories in two
dimensions, the components of the energy-momentum tensor generating
antiholomorphic reparametrizations vanish in $Q$-cohomology.  Thus,
the twisted theories are holomorphic.  The $(2,0)$ theory reduces to
theories of these kinds by appropriate compactification on $C$ or $M$,
so we expect that the twisted $(2,0)$ theory becomes topological along
$M$ and holomorphic along $C$ if we think of $Q$ as a BRST operator.

This expectation is backed up by the existence of an analogous twist
in four dimen\-sions, studied by Kapustin in \cite{Kapustin:2006hi}.
Kapustin's twist can be applied to any $\CN = 2$ gauge theory with
nonanomalous $\SU(2) \times \U(1)$ R-symmetry group, formulated on the
product $\Sigma \times C$ of Riemann surfaces.  Such theories are
superconformal.  We write the $\U(1)$ factor of \linebreak $\SU(2)
\times \U(1)$ as $\U(1)_\CR$, to distinguish from a maximal torus of
the $\SU(2)$ subgroup which we denote by $\U(1)_R$.  The holonomy
group of $\Sigma \times C$ is $\U(1)_\Sigma \times \U(1)_C$.  One
replaces $\U(1)_\Sigma$ \linebreak with the diagonal subgroup of
$\U(1)_\Sigma \times \U(1)_R$.  Similarly, one twists $\U(1)_C$ with
$\U(1)_\CR$.  After twisting, two of the eight supercharges, $Q_\ell$
and $Q_r$, are scalars.  For the BRST operator \linebreak one takes a
linear combination
\begin{equation}
  Q = Q_\ell + tQ_r
\end{equation}
with $t \in \C^\times$.  Then, the twisted theory is topological along
$\Sigma$ and holomorphic along $C$.  For $t = 0$ or $\infty$, it is
holomorphic both on $\Sigma$ and $C$.

The relevance of Kapustin's twist to the present story is that if one
takes $M = \Sigma \times \Sigma'$, with $\Sigma'$ another Riemann
surface, and compactifies the twisted $(2,0)$ theory on $\Sigma'$, one
can obtain an $\CN = 2$ superconformal theory on $\Sigma \times C$.
The complex structure of $\Sigma'$ determines the complexified gauge
couplings which combine the gauge couplings and the $\theta$-angles.
If the twisted theory has the claimed quasi-topological property, then
$\CN = 2$ theories so obtained have to be twisted in such a way that
they are topological along $\Sigma$ and holomorphic along $C$, and
moreover independent of the couplings.  Kapustin's twist does produce
theories with the desired properties, and it is essentially the only
such twist.  The $\U(1)_\CR$ symmetry of the $(2,0)$ theory is the
$\U(1)_\CR$ group used in Kapustin's twist, while a maximal torus of
$\SU(2)_R$ is the $\U(1)_R$ group.

Having defined the twisted $(2,0)$ theory on $M \times C$, we now want
to turn on the $\Omega$-deformation.  Let us first recall the standard
formulation in four dimensions.

The $\Omega$-deformation of an $\CN = 2$ gauge theory on a
four-manifold $M$ can be considered when $M$ admits a $\U(1)$
isometry.  Denote by $V$ the vector field generating the isometry.
Then, the procedure is roughly to replace the adjoint scalar $\phi$ in
the vector multiplet as
\begin{equation}
  \phi \to \phi + \epsilon V^\mu D_\mu,
\end{equation}
where $\epsilon$ is a complex parameter and $D = d + A$ is the
covariant derivative.  Since $\phi$ is replaced \linebreak by a
differential operator, this is not a change of variables but a
deformation of the theory.

Like the undeformed case, the $\Omega$-deformed theory is topological
after twisting, but with respect to a different supercharge.  The
twisted supercharges are a scalar $Q$, a one-form $Q_\mu$, and a
self-dual two-form $Q_{\mu\nu}$.  Of these, at least $Q$ and $V^\mu
Q_\mu$ are unbroken in the undeformed theory.  The
$\Omega$-deformation preserves the linear combination
\begin{equation}
  \Qt = Q + \epsilon V^\mu Q_\mu.
\end{equation}
To make the $\Omega$-deformed theory topological one takes $\Qt$ as
the BRST operator.

We will consider the case $M = \R^4$.  In this case we can deform the
theory with two commuting $\U(1)$ isometries, rotating two orthogonal
two-planes in $\R^4$.  To do this, we just replace $\epsilon V$ in the
above formulas by $\epsilon_1 V_1 + \epsilon_2 V_2$, where $V_1$ and
$V_2$ are the Killing vector fields for the rotations, and
$\epsilon_1$ and $\epsilon_2$ are the corresponding parameters.

As stressed already, we will assume that the $\Omega$-deformation of a
twisted $\CN = 2$ gauge theory lifts to a deformation of the
underlying twisted $(2,0)$ theory.  Later we will rephrase our
assumption in somewhat different terms.

To better understand the effect of the $\Omega$-deformation to the
twisted $(2,0)$ theory, we \linebreak want to compactify the theory to
lower dimensions where a Lagrangian description is available.  To this
end we equip $\R^4$ with the metric for $D_1 \times D_2$, the product
of two cigars whose \linebreak radii $\rho_1$ and $\rho_2$ we will
take to be small.  It turns out that the $\Omega$-deformation has a
particularly nice description in this setup, thanks to the following
property of the cigar geometry observed by Nekrasov and Witten
\cite{Nekrasov:2010ka}: the $\Omega$-deformation on a cigar can be
canceled by a change of variables everywhere except in the curved
region near the tip.  Here we briefly explain why this is true.

Consider an $\CN = 2$ gauge theory on the product of a two-manifold
and a flat cylinder of radius $\rho$.  We turn on the
$\Omega$-deformation using the rotation about the axis of the
cylinder.  Since the $\U(1)_R$ symmetry rotates $\epsilon$ by a phase,
we can assume that $\epsilon$ is real.  Then, the $\Omega$-deformation
is realized by the substitution
\begin{equation}
  \label{A4}
  A_4 \to A_4 + \epsilon \rho D_1,
\end{equation}
where $A_4$ is defined by writing $\phi = A_4 + iA_5$ using
antihermitian adjoint-valued one-forms $A_4$ and $A_5$, and $x^1 \sim
x^1 + 2\pi\rho$ is the coordinate in the circle direction of the
cylinder.

Let us combine this with the rescaling of the radius
\begin{equation}
  \label{rho}
  \rho \to \rho\sqrt{1 + \epsilon^2 \rho^2},
\end{equation}
and introduce the rescaled coordinate $\xh^1 = x^1/\sqrt{1 +
  \epsilon^2 \rho^2}$ of the same periodicity as before.  Also, make
the change of variables $\Ah_4 = A_4/\sqrt{1 + \epsilon^2 \rho^2}$ and
$\Ah_1 = A_{\hat 1} + \epsilon\rho \Ah_4$, and define $\Dh_1 =
\del/\del\xh^1 + \Ah_1$.  In terms of these, the effect of the
combined operations \eqref{A4} and \eqref{rho} is to replace
\begin{equation}
  \label{A4h}
  A_4
  \to \frac{1}{\sqrt{1 + \epsilon^2 \rho^2}} (\Ah_4 + \epsilon \rho \Dh_1),
\end{equation}
while we have
\begin{equation}
  \label{D1}
  D_1
  = \frac{1}{\sqrt{1 + \epsilon^2 \rho^2}} (\Dh_1 - \epsilon \rho \Ah_4).
\end{equation}

The point is that the two substitutions \eqref{A4h} and \eqref{D1}
together can be thought of as a rotation in the $A_4$-$D_1$ plane, and
under such a rotation (followed by the corresponding rotation of
fermions) the action of the undeformed theory is invariant.  In other
words, the $\Omega$-deformation on a cigar can be undone by a change
of variables and a rescaling of the radius.  The undeformed theory so
obtained has the coupling constant rescaled as
\begin{equation}
  \label{e}
  e^2 \to \frac{e^2}{\sqrt{1 + \epsilon^2\rho^2}},
\end{equation}
since the volume form contains the factor $dx^1 = \sqrt{1 +
  \epsilon^2\rho^2} d\xh^1$.

One can replace the cylinder by a cigar, and apply the same argument
in the region where the cigar looks like a flat cylinder.  Then one
sees that the $\Omega$-deformation can be canceled away from the tip
of the cigar, by a change of variables.  This was the insight of
Nekrasov and Witten.

We can readily extend the above argument to the situation in which the
theory is for-\linebreak mulated on $D_1 \times D_2$.  Taking the
coordinates in the circle directions to be $x^1 \sim x^1 + 2\pi\rho_1$
and $x^2 \sim x^2 + 2\pi\rho_2$, away from the tips of the cigar the
$\Omega$-deformation is given by the substitution
\begin{equation}
  A_4 \to A_4 + \epsilon_1 \rho_1  D_1 + \epsilon_2 \rho_2 D_2.
\end{equation}
This is the same operation as the $\Omega$-deformation \eqref{A4} for
a single cylinder if we set\linebreak $(\epsilon\rho)^2 = (\epsilon_1
\rho_1)^2 + (\epsilon_2 \rho_2)^2$ and rotate the $x^1$-$x^2$ plane by
angle $\tan^{-1}(\epsilon_1 \rho_1/\epsilon_2 \rho_2)$.  The rotation
does not change the form of the action or the standard flat metric, so
the $\Omega$-deformation can be canceled%
\footnote{The argument is presented here for the case when
  $\epsilon_1$ and $\epsilon_2$ are both real.  The general case can
  be treated as follows.  The general $\Omega$-deformation is given by
  the substitutions $A_4 \to A_4 + \Re \epsilon_1 \rho_1 D_1 + \Re
  \epsilon_2 \rho_2 D_2$ and $A_5 \to A_5 + \Im \epsilon_1 \rho_1 D_1
  + \Im \epsilon_2 \rho_2 D_2$.  If $\theta_1 = \arg \epsilon_1$ and
  $\theta_2 = \arg \epsilon_2$ are equal, one can make $\epsilon_1$
  and $\epsilon_2$ both real by a $\U(1)_R$ rotation.  So suppose
  $\theta_1 \neq \theta_2$.  The $\Omega$-deformation can be canceled
  if these substitutions can be put into the form $A_4 \to A_4 +
  \epsilon_1' \rho_1 D_1$ and $A_5 \to A_5 + \epsilon_2' \rho_2 D_2$
  for some $\epsilon_1'$, $\epsilon_2' \in \R$, by rotating the
  $x^1$-$x^2$ plane.  This requires $\Re \epsilon_1\rho_1/\Re
  \epsilon_2\rho_2 = -\Im \epsilon_2\rho_2/\Im \epsilon_1\rho_1$, or
  $\sin(2\theta_1)/\sin(2\theta_2) = - |\epsilon_2|^2
  \rho_2^2/|\epsilon_1|^2 \rho_1^2$.  Since the function $f(\alpha) =
  \sin(2\theta_1 + \alpha)/\sin(2\theta_2 + \alpha)$ ranges from
  $-\infty$ to $+\infty$ as $\alpha$ varies, it is possible to satisfy
  this condition by a $\U(1)_R$ rotation.}
by a change of variables together with a rescaling of the metric and
the coupling by a factor of $1/\sqrt{1 + \epsilon^2\rho^2}$.

Therefore, the $\Omega$-deformed action $\St$, written in nonstandard
variables, is equal to the undeformed action $S$ with the rescaled
metric and coupling, plus terms that vanish on the flat cylinder
region.  These terms are $\Qt$-invariant, as both $\St$ and $S$ are.
To put it differently, the effect of the $\Omega$-deformation on $D_1
\times D_2$ is essentially to insert $\Qt$-invariant operators
supported in the neighborhood of $\{0\} \times D_2$ and $D_1 \times
\{0\}$, where $0 \in D_1$ or $D_2$ is the tip of either cigar.

In the limit where the radii of the cigars are sent to zero, $\Qt$
reduces to $Q$ and the curvature localizes to the tips.  Thus the
inserted operators are well approximated by $Q$-invariant operators
supported at the tips, and the approximation becomes arbitrarily good
as the radii go to zero.  Reversing the logic, for twisted $\CN = 2$
gauge theories on $\R^4$, we may define the $\Omega$-deformation by
the insertion of these $Q$-invariant operators.  For this definition
coincides with the standard one after $\R^4$ is deformed to $D_1
\times D_2$, the metric and the coupling are rescaled, and the zero
radii limit is taken, the operations none of which affects the
topological field theory.  (The Kaluza-Klein modes discarded in taking
the zero radii limit are not important, as the momenta are identically
zero in $Q$-cohomology.)

Then, the assumption we really want to make is that these
$Q$-invariant operators lift to $Q$-invariant operators in the twisted
$(2,0)$ theory on $D_1 \times D_2 \times C$, supported at the tips of
the cigars, $\{0\} \times D_2 \times C$ and $D_1 \times \{0\} \times
C$.  The insertion of the lifted operators may be regarded as a
definition of the $\Omega$-deformation of the twisted $(2,0)$ theory
on $\R^4 \times C$.

\section{Down to four dimensions}
\label{4d}

The motivation for taking the small radii limit was to compactify the
$(2,0)$ theory to lower dimensions where we understand things better.
So let us look at the effective description of the theory.

Compactification on the circle fibers of the cigars gives a theory on
the four-manifold with a corner $L_1 \times L_2 \times C$, where $L_1$
and $L_2$ are half-lines.  The tips of the cigars correspond to the
boundaries $\{0\} \times L_2 \times C$ and $L_1 \times \{0\} \times
C$.  The $(2,0)$ theory compactified on a torus is $\CN = 4$ super
Yang-Mills theory whose complexified gauge coupling $\tau =
\theta/2\pi + 4\pi i/e^2$ is given by the complex structure modulus of
the torus.  In our case, away from the tips we are compactifying the
theory on a rectangular torus with sides of length $2\pi\rho_1$ and
$2\pi\rho_2$.  Thus, the bulk theory is $\CN = 4$ super Yang-Mills
theory with $\tau = i\rho_1/\rho_2$.  The gauge group is the compact
Lie group $G$ associated with the simply-laced Lie algebra $\gf$
specifying the type of the $(2,0)$ theory.

The twisting of the $(2,0)$ theory reduces to the Kapustin twist,
which makes the theory topological along $L_1 \times L_2$ and
holomorphic along $C$.  The compactified theory has a fixed value of
$t$ as the twisted $(2,0)$ theory carries only one supercharge.  The
precise value is inessential, however; the $\U(1)_R$ symmetry rotates
$Q_\ell$ and $Q_r$ by opposite phases, so one can set $t$ to any value
except $0$ or $\infty$ by the complexified $\U(1)_R$-action applied in
six dimensions, without changing the $Q$-cohomology.  It is convenient
to set $t = i$.

Turning on the $\Omega$-deformation in the twisted $(2,0)$ theory
means inserting $Q$-invariant operators placed at the tips of the
cigars.  These operators descend to $Q$-invariant boundary couplings
in the $\CN = 4$ theory.  Eventually we will deduce what the boundary
couplings are, but before doing that, we want to identify the boundary
conditions in the absence of the $\Omega$-deformation.  For this
purpose it is helpful to resort to brane constructions.

We will assume that the cotangent bundle $T^*C$ of $C$ admits a
complete Calabi-Yau metric.  This is the case if $C$ is a sphere or a
torus, for example.  For $C = S^2$, one can endow $T^*C$ with the
Eguchi-Hanson metric.

First we focus on the boundary created at the tip of $D_1$, so let us
forget about the cap of $D_2$, replacing it with a cylinder $\Dt_2$.
We denote the axis of $\Dt_2$ by $\Lt_2 = \R$.  We consider $N$
M5-branes wrapped on $D_1 \times \Dt_2 \times \{0\} \times C$ in
M-theory on $T^*D_1 \times \Dt_2 \times \R \times T^*C$.  If $T^*D_1$
and $T^*C$ are endowed with Calabi-Yau metrics, so that some of the
supersymmetries are preserved, then the low-energy dynamics of the
M5-branes realizes the $(2,0)$ theory of type $A_{N-1}$ on $D_1 \times
\Dt_2 \times C$.  With $D_1$ being a cigar, we can choose $T^*D_1$ to
be a Taub-NUT space \linebreak $\TN$.  This space is a circle
fibration over $\R^3$ such that the radius of the circle shrinks to
zero \linebreak at the origin and approaches a finite asymptotic value
at infinity, so the fibers over each radial direction make up a cigar.
To embed $D_1$ in $\TN$, we pick a direction $L_1$ and identify $D_1$
with the fibers over $L_1$.  For clarity we write $\Lt_1 = \R$ and
$\R^3 = \Lt_1 \times \R^2$, with $L_1 \subset \Lt_1$.

We compactify this system on the $S^1$ of $\Dt_2 = \Lt_2 \times S^1$.
This gives $N$ D4-branes wrapped on $D_1 \times \Lt_2 \times \{0\}
\times C$ in Type IIA string theory on $\TN \times \Lt_2 \times
\R \times T^*C$.  We still want to compactify the system on the circle
of $D_1$.  So we perform a $T$-duality to unwrap the branes from the
circle, and take the radius of the dual circle $\Sh^1$ to be large.
At first sight, the $T$-duality may appear to replace $\TN$ by the
dual fibration with fiber $\Sh^1$, and turns the D4-branes to
D3-branes.  This is actually not the case.  Rather, it produces a
D3-NS5 system in Type IIB string theory on
\begin{equation}
  \label{ST}
  \Sh^1 \times \Lt_1 \times \R^2 \times \Lt_2 \times \R \times T^*C,
\end{equation}
with $N$ D3-branes ending on a single NS5-brane \cite{Townsend:1995kk,
  Gregory:1997te}.  The support of the D3-branes is
\begin{equation}
  \{p\} \times L_1 \times \{0\} \times \Lt_2 \times \{0\} \times C,
\end{equation}
while the NS5-brane is wrapped on
\begin{equation}
  \label{NS5}
  \{p\} \times \{0\} \times \{0\} \times \Lt_2 \times \R \times T^*C,
\end{equation}
where $p$ is a point of $\Sh^1$.  The low-energy dynamics of the
D3-branes is described by $\CN = 4$ super Yang-Mills theory on $L_1
\times \Lt_2 \times C$.  We have a half-BPS boundary condition on the
boundary located at $0 \in L_1$, where the D3-branes end on the
NS5-brane.

Only the local geometry matters to the boundary condition.  Thus, if
we ignore the curvature of $C$ for simplicity, the above D3-NS5 system
leads to the same boundary condition as that for a system of $N$ flat
D3-branes ending on a single flat NS5-brane, with the D3- and
NS5-branes sharing three spacetime directions.  The $\CN = 4$ theory
on the D3-branes has six scalars coming from the fluctuations in the
six normal directions.  The rotations in the normal plane give rise to
the R-symmetry group $\SO(6)$, but the presence of the NS5-brane
breaks it to $\SO(3)_X \times \SO(3)_Y$, the product of the rotation
groups of the three-planes tangent and normal to the NS5-brane.  The
breaking of $\SO(6)$ divides the scalars into two triplets, $\vec X$
and $\vec Y$, transforming as a vector under $\SO(3)_X$ and
$\SO(3)_Y$, respectively.  The NS5-brane imposes Neumann boundary
conditions on $\vec X$ and the gauge field $A$; writing $x$ for the
coordinate of $L_1$, on the boundary we have
\begin{equation}
  \label{NBC}
  D_x \vec X = F_{x\mu} = 0,
\end{equation}
where $F$ is the curvature of $A$.  Fluctuations normal to the
NS5-brane must vanish on the boundary, so $\vec Y$ obey Dirichlet
boundary conditions:
\begin{equation}
  \label{Y=0}
  \vec Y = 0.
\end{equation}
For the fermions, the boundary conditions set half of them to zero.
Lastly, requiring that the boundary conditions themselves be invariant
under the would-be unbroken supersymmetries imposes further conditions
on the fermions.

Next, we look at the boundary coming from the tip of $D_2$, this time
replacing $D_1$ with a cylinder $\Dt_1$ with axis $\Lt_1 = \R$.  So we
start with a system of $N$ M5-branes wrapped on $\Dt_1 \times D_2
\times \{0\} \times C \subset \Dt_1 \times \TN' \times \R \times
T^*C$, with $\TN'$ another Taub-NUT space, and first compactify it on
$D_2$.  We embed $D_2$ in $\TN'$ as the fibers over a radial direction
$L_2$ in the base $\R^3 = \R^2 \times \Lt_2$.  Then, upon
compactification, the M5-branes become D4-branes wrapped on $\Dt_1
\times \{0\} \times L_2 \times \{0\} \times C$ in the spacetime $\Dt_1
\times \R^2 \times \Lt_2 \times \R \times T^*C$.  In addition, the
compactification creates a D6-brane supported at the origin of the
base \cite{Townsend:1995kk, Gregory:1997te}, that is, on $\Dt_1 \times
\{0\} \times \{0\} \times \R \times T^*C$.  After that, we perform a
$T$-duality on the $S^1$ of $\Dt_1 = S^1 \times \Lt_1$.  The end
result is a D3-D5 system in the spacetime \eqref{ST}, with $N$
D3-branes ending on a D5-brane.  The D3-branes are supported on
\begin{equation}
  \{p\} \times \Lt_1 \times \{0\} \times L_2 \times \{0\} \times C
\end{equation}
and the D5-brane is on
\begin{equation}
  \label{D5}
  \{p\} \times \Lt_1 \times \{0\} \times \{0\} \times \R \times T^*C.
\end{equation}

The D3-D5 system we have arrived at is the $S$-dual of the D3-NS5
system.  This is consistent with the fact that the $S$-duality of $\CN
= 4$ super Yang-Mills theory is realized in six dimensions as the
modular transformations acting on the torus on which the $(2,0)$
theory is compactified.  When the torus is rectangular, interchanging
the two sides gives the duality $\tau \to -1/\tau$.  In the D3-D5
construction the roles of $D_1$ and $D_2$ are switched compared to the
D3-NS5 construction, so in effect we have applied $S$-duality.

At low energies the dynamics of the D3-branes is described by $\CN =
4$ super Yang-Mills \linebreak theory on $\Lt_1 \times L_2 \times C$,
with the $S$-dual half-BPS boundary condition.  The presence of the
D5-brane breaks the R-symmetry group $\SO(6)$ to $\SO(3)_{X'} \times
\SO(3)_{Y'}$, and divides the scalars into two triplets $\vec X'$ and
$\vec Y'$.  Writing $y$ for the coordinate of $L_2$, the D3-D5
bound\-ary conditions are such that near the boundary $\vec X'$ are
approximated by a solution of the Nahm equations
\begin{equation}
  \label{Nahm}
  \frac{DX'_i}{Dy} + \epsilon_{ijk} [X'_i, X'_j] = 0,
\end{equation}
with the particular singular behavior
\begin{equation}
  \label{NP}
  \vec{X'} = \frac{\vec{t}}{y} + \dotsb.
\end{equation}
Here $\vec t$ is a triplet of elements in $\gf$ giving a principal
$\suf(2)$ embedding with the standard commutation relations $[t_i,
t_j] = \epsilon_{ijk} t_k$, and the ellipsis refers to terms less
singular than $1/y$ \cite{Gaiotto:2008sa}.  We can gauge away $A_y$ at
$y = 0$.  The other components of the gauge field obey Dirichlet
boundary conditions,%
\footnote{If $C$ is curved, the boundary value of the gauge field must
  actually be related to the Riemannian connection on the boundary in
  a specific manner \cite{Witten:2011zz}.  This modification does not
  affect our analysis.}
as do the scalars $\vec{Y}'$:
\begin{equation}
  \label{DBC}
  A = \vec Y' = 0.
\end{equation}
The boundary condition sets to zero half of the fermions, possibly
different from the previous half.

Combining the two descriptions of the compactification, we conclude
that the $(2,0)$ theory of type $A_{N-1}$ compactified on the circle
fibers of $D_1 \times D_2$ is realized by a D3-D5-NS5 system in Type
IIB string theory in the spacetime \eqref{ST}, with $N$ D3-branes
ending on a D5-brane in one direction and on an NS5-brane in another.
The D3-branes are supported~on
\begin{equation}
  \{p\} \times L_1 \times \{0\} \times L_2 \times \{0\} \times C,
\end{equation}
and the NS5- and D5-branes are wrapped on the same submanifolds
\eqref{NS5} and \eqref{D5} as before.  Notice that the directions
normal to the D3- and NS5-branes are the same as those normal to the
D3- and D5-branes.  Thus $\vec X = \vec X'$ and $\vec Y = \vec Y'$.
Otherwise, the D3-NS5 and D3-D5 boundary conditions would not be
compatible.

The boundary conditions derived from the D3-D5-NS5 system should
preserve the su\-persymmetry of the twisted theory.  One may feel that
this point is somewhat obscured in the above construction, since we
looked at the two boundaries separately.  Nonetheless, this must be
true, as the following argument shows.  Ideally, one could start with
a system of M5-\linebreak branes wrapped on $D_1 \times D_2 \times C
\subset V \times T^*C$, where $V$ is a seven-manifold of $G_2$
holonomy in \linebreak which $D_1 \times D_2$ is embedded as a
supersymmetric cycle.  This system has only one unbroken supersymmetry
in general, and it is precisely the one that is preserved by the
particular twist we applied to the $(2,0)$ theory
\cite{Bershadsky:1995qy}.  Instead of picking $V$, we started with
$\TN \times \Dt_2 \times \R$ \linebreak or $\Dt_1 \times \TN' \times
\R$ pretending that $D_2$ or $D_1$ was a cylinder.  These are
certainly $G_2$ manifolds, their holonomy being $\SU(2) \subset G_2$.
Therefore, the D3-NS5 and D3-D5 systems both pre\-serve the
supersymmetry of the twisted theory, hence so does the total D3-D5-NS5
system.

There is a question about which scalar on the D3-branes corresponds to
which in the twisted theory.  We can determine the correspondence from
the fact that the D3-NS5 boundary condition admits a generalization to
the case with nonvanishing $\theta$-angle \cite{Gaiotto:2008sa}.  For
$\theta \neq 0$,\linebreak the simple Neumann boundary conditions
\eqref{NBC} are modified due to the boundary coupling by a
Chern-Simons term, constructed from the complex gauge field given by a
linear combination of $A$ and $\vec X$ \cite{Witten:2011zz}.  For such
a boundary coupling to make sense when $C$ is curved, $\vec X$ must be
twisted to a one-form of the type $X_y dy + X_z dz + X_\zb d\zb$,
where $z$ is a holomorphic coordinate on $C$.  This allows us to
identify, up to $\SO(3)_X$ rotations and rescalings,
\begin{equation}
  \begin{aligned}
    X_z &= X_1 + iX_2, \\
    X_\zb &= X_1 - iX_2, \\
    X_y &= X_3.
  \end{aligned}
\end{equation}
As for $\vec Y$, after the twisting one of them becomes a one-form
$Y_x dx$ and the others remain to be scalars.

\section{Turning on the \texorpdfstring{$\mathbf\Omega$}{Omega}-deformation}
\label{WZW}

Now we turn on the $\Omega$-deformation, and try to determine the
induced $Q$-invariant boundary couplings and the dynamical degrees of
freedom that emerge on ``the boundary of the boundaries.''  Since the
boundary couplings for the two boundaries should be related by
$S$-duality, we will only consider the one for the D3-NS5 boundary.
For simplicity, we will assume that $C$ has no boundary, and write $L
\times C$ for $\{0\} \times L_2 \times C$ with $L$ a half-line.
 
The boundary coupling is constructed out of the gauge field $A$ and
the adjoint-valued one-form $X$, and half of the fermions.%
\footnote{Adding boundary couplings modifies the boundary conditions
  \eqref{NBC} and \eqref{Y=0} which are appropriate for the standard
  $\CN = 4$ super Yang-Mills action.  As we will see shortly, however,
  the added terms are independent of the gauge coupling and so one can
  take the weak coupling limit where the modified conditions reduce to
  the original ones.}
It must satisfy two criteria derived from the quasi-topological nature
of the twisted $(2,0)$ theory.

First, the compactified theory must be holomorphic along $C$, and the
bulk theory already has this property, so the boundary coupling must
also have the same property.

Second, the compactified theory must be independent of the coupling
constant $e^2$ given by the ratio of the radii of the cigars, and the
bulk theory is already independent, so the boundary coupling must not
introduce a coupling dependence.

The second condition is not as weak as it may sound, because the bulk
theory realizes the coupling independence in an interesting manner
\cite{Kapustin:2006hi}.  Some terms of the bulk action can\-not be
written in a $Q$-exact form, apparently leading to a coupling
dependence.  Still, they fail to be $Q$-exact by terms quadratic in
fermions.  So the coupling dependence can actually be absorbed
entirely by a rescaling of the fermions.  This rescaling makes it hard
for the boundary coupling to be supersymmetric (and not $Q$-exact).
Typically, a supersymmetric action contains purely bosonic piece
related by supersymmetry to pieces involving fermions, with relative
coefficients independent of the coupling.  If the bosonic part is
coupling inde-\linebreak pendent, then the rescaling introduces a
coupling dependence to the fermionic part.

There is a natural candidate satisfying the two criteria: a
Chern-Simons term with level independent of the coupling.  In fact,
the twisted $\CN = 4$ super Yang-Mills theory has a $Q$-invariant
complex gauge field%
%
%
\begin{equation}
  \CA = (A_x + iY_x) dx + (A_y + iX_y) dy + X_z dz + A_\zb d\zb,
\end{equation}
from which one can construct the Chern-Simons action
\begin{equation}
  \label{CS}
  S_{\ChS}[\CA]
  = \frac{1}{4\pi i} \int_{L \times C}
    \Tr\Bigl(\CA \wedge d\CA + \frac{2}{3} \CA \wedge \CA \wedge \CA\Bigr)
\end{equation}
for the complexified gauge group $G_\C$.  Here $\Tr$ denotes the
Killing form divided by twice the dual Coxeter number $h^\vee$ of $G$;
for $G = \SU(N)$, it equals the trace in the $N$-dimensional
representation.

However, the Chern-Simons term cannot be all that is present, for it
does not lead to the correct boundary conditions.  In terms of $\CA$,
the Nahm pole boundary conditions~\eqref{NP}~read
\begin{equation}
  \label{NP-A}
  \begin{aligned}
    \CA_z  &= \frac{t_+}{y} + \dotsb, \\
    \CA_\zb &= 0, \\
    \CA_y  &= \frac{it_3}{y} + \dotsb,
  \end{aligned}
\end{equation}
where $t_+ = t_1 + it_2$.  Under variations $\delta \CA$, the
Chern-Simons action changes by
\begin{equation}
  \delta S_{\ChS}[\CA]
  = \frac{1}{4\pi i} \int_{L \times C}
    \Tr\bigl(2\delta\CA \wedge \CF
    - \del_y(\CA_z \delta\CA_\zb - \CA_\zb \delta\CA_z) dy \wedge dz \wedge d\zb\bigr).
\end{equation}
The variation is not necessarily zero even if the bulk equations of
motion $\CF = 0$ is satisfied and the boundary conditions \eqref{NP-A}
are imposed, since $\CA_z \delta\CA_\zb$ may not go to zero as $y \to
0$ due to the presence of singular terms in $\CA_z$.  To fix the
problem, we add the term
\begin{equation}
  \label{CSB}
  S_\del[\CA]
  = \frac{1}{4\pi i} \int_{L \times C}
    \Tr \del_y(\CA_z \CA_\zb) dy \wedge dz \wedge d\zb,
\end{equation}
which would be a boundary term if $\CA_z$ were nonsingular.  The
variation then vanishes up to the bulk equations of motion,
provided that $\delta \CA$ respects the boundary
conditions~\eqref{NP-A} and is regular at $y = 0$.

We therefore propose that the boundary coupling at $y = 0$ is given by
a Chern-Simons term \eqref{CS} supplemented with the boundary
term \eqref{CSB}:
\begin{equation}
  kS_{\ChS + \del}.
\end{equation}

At this point one may object that there can be other possibilities for
the boundary coupling.  In fact, any gauge-invariant boundary terms
whose contributions to the energy-momentum tensor are $Q$-exact
(except the component $T_{zz}$) seem suitable as long as they do not
introduce a coupling dependence.  However, those terms are not very
interesting to us; when placed on a manifold with boundary, they do
not induce dynamical degrees of freedom localized on the boundary.
The emergence of such boundary degrees of freedom is a consequence of
the breaking of gauge invariance by the boundary.  We need something
that is not gauge invariant, and yet gives a gauge-invariant quantity
when integrated over a manifold without boundary.

For Chern-Simons theory, the boundary degrees of freedom are described
by a WZW model \cite{Witten:1988hf, Moore:1989yh, Elitzur:1989nr}.
This can be seen in the path integral formalism as follows
\cite{Ogura:1989gn, Carlip:1991zm}.  Under gauge transformation
\begin{equation}
  \CA \to \CA^g = g^{-1} \CA g + g^{-1} dg,
\end{equation}
the Chern-Simons action on a three-manifold $V$ with boundary
transforms as
\begin{equation}
  \label{CS-gWZW}
  S_{\ChS + \del}[\CA] = S_{\ChS + \del}[\CA^g] + S_{\gWZW}[g, \CA^g],
\end{equation}
where $S_{\gWZW}$ is the following gauged WZW action:
\begin{equation}        
  \label{gWZW}
  S_{\gWZW}[g, A]
  = \frac{1}{4\pi i} \int_{\del V}
    \Tr\bigl((g^{-1} \del g - 2A^{1,0}) \wedge g^{-1} \delb g\bigr)
    + \frac{1}{12\pi i} \int_V \Tr(g^{-1}dg)^3.
\end{equation}
Consider the path integral in a neighborhood $W$ of the boundary.  The
gauge inequivalent configurations of $\CA$ in $W$ form a space
$\CM_W$, the moduli space of $G_\C$-connections over $W$.  If one
makes a gauge choice $\CAh$ at each point of $\CM_W$ and defines the
Faddeev-Popov determinant $\Delta$ by
\begin{equation}
  \label{FP}
  1 = \Delta(\CA)
      \int \! \CD g \, \delta(\CA^g - \CAh),
\end{equation}
then the path integral measure can be written as
\begin{equation}
  \label{CSWZW}
  \int \!\CD\CA \exp(-kS_{\ChS+\del}[\CA])
  = \int_{\CM_W} \!\!\! \Delta(\CAh) \! \int \! \CD g
    \exp\bigl(-k(S_{\ChS+\del}[\CAh] +S_{\gWZW}[g, \CAh])\bigr).
\end{equation}
The gauge degrees of freedom are therefore converted on the boundary
to dynamical ones, described by the WZW action coupled to the
background gauge field $\CAh$.

The action $S_\gWZW[g, \CAh]$ coincides with the standard WZW action
if one chooses $\CAh_z = 0$ on the boundary, which is always possible
as the integrability condition $\del_A^2 = 0$ is trivially satisfied
in complex dimension one.  Even so, the emergent boundary theory is
generally not the ordinary WZW model due to constraints imposed on
gauge transformations by the boundary conditions.  A more precise way
of saying this is that since the boundary conditions reduce the space
of connections and hence the space of boundary gauge transformations
required for gauge fixing, the path integral in the definition
\eqref{FP} of the Faddeev-Popov determinant should be performed over
this reduced space.  If the boundary conditions impose first-class
constraints, the resulting theory will be a gauged version of the WZW
model.

Let us apply the above considerations to our case.  In order to gauge
away $\CA_z$ near $y = 0$, we must allow gauge transformations to be
singular at $y = 0$.  So we pick some small $\delta > 0$ and integrate
over maps $g\colon [0, \delta) \times C \to G_\C$, including singular
ones.%
\footnote{The formula \eqref{gWZW} still makes sense if we extend $g$
  to $L \times C$ and rewrite everything as an integral over $L \times
  C$ using Stokes' theorem.  However, here we must be more careful
  about what we mean by the path integral over $g$, since the
  integrand $\exp(-S_\gWZW)$ is unbounded due to the noncompactness of
  $G_\C$.  (It would be fine to include singular maps otherwise;
  configurations with diverging kinetic energy simply do not
  contribute.)  The same issue actually arises at the level of
  Chern-Simons theory, so the method developed in \cite{Witten:2010cx}
  may be adapted to provide a proper path integral definition.  Our
  considerations do not depend on whether or not the path integral
  formalism is applicable.}
Not all of these maps are equally important, though.  The maps that
are really relevant are those for which $\CA^g = \CAh$.  These maps
are singular, but we can make a singular gauge transformation so that
many of them become regular at $y = 0$ and admit a simple
interpretation as boundary degrees of freedom.

Let $\{t_a\}$ be a basis of $\gf_\C$ such that $[it_3, t_a] = s_a
t_a$, and split $\gf_\C$ into the subalgebras $\gf_+$, $\gf_0$, and
$\gf_-$ of $s_a$ positive, zero, and negative.  Conjugation by $g_y =
\exp(-it_3 \ln y)$ acts on $t_a$ as multiplication by $y^{-s_a}$.
Thus, after the gauge transformation by $g_y$ the boundary condition
for $\CA_z$ becomes
\begin{equation}
  \label{NP-A2}
  \CA_z   = t_+ + \sum_{t_a \in \gf_0 \oplus \gf_-} y^{s_a} f_z^a t_a,
\end{equation}
where the coefficient functions $f_z^a$ are less singular than $1/y$.
(For $\vec t$ giving a principal embedding, the $s_a$ are integers and
thus $y^{s_a} f(y) \to 0$ as $y \to 0$ if $s_a > 0$ and $f(y)$ is less
singular than $1/y$.)  Since $g_y$ leaves invariant the action
$S_{\ChS + \del}[\CA]$, and we want to consider singular gauge
transformations anyway, we could as well impose these boundary
conditions from the beginning.

The path integral decomposes into different sectors classified by the
behavior of the fields at $y = 0$.  If we restrict our attention to
the sector in which $\CA_z$ are regular, then gauge transformations
setting $\CA_z$ to zero are also regular, as we desired.  From now on
we will focus on this sector.

The left action $g \mapsto hg$ by maps $h(z)$, holomorphic along $C$
and constant along $L$, \linebreak would be a symmetry of the boundary
theory if the theory were the ordinary WZW model.  The usual story is
that this symmetry implies the existence of an affine Kac-Moody
alge-\linebreak bra $\gfh$ of level $k$ in the chiral algebra.  (The
antiholomorphic counterpart of this algebra is trivial in
$Q$-cohomology.)  Here we get a smaller algebra because of the
boundary condi-\linebreak tions.  Indeed, the gauge-fixing condition
$\CA_z^g = 0$ relates the affine currents $J = J_z dz$ to the boundary
value of $\CA_z$:
\begin{equation}
  J_z = -k \del_z g g^{-1}|_{y = 0} = k\CA_z|_{y = 0}.
\end{equation}
Comparing this formula with the boundary condition \eqref{NP-A2}, we
see that the affine currents obey
\begin{equation}
  \label{J-constraint}
  \sum_{t_a \in \gf_+} J^a t_a - kt_+ = 0.
\end{equation}
This equation encodes first-class constraints; the left-hand side
generates the gauge transformations by the subgroup $G_-$ of $G_\C$
with Lie algebra $\gf_-$.  Therefore, the boundary degrees of freedom
are described by a $G_-$-gauged WZW model.

To summarize, we have argued that the boundary coupling on $\{0\}
\times L_2 \times C$ is given by \linebreak a Chern-Simons term with
the Nahm pole boundary conditions \eqref{NP-A}, which induce
dynam-\linebreak ical degrees of freedom on $\{0\} \times \{0\} \times
C$ described by the above gauged WZW model.  On the other boundary
$L_1 \times \{0\} \times C$, we have dual boundary coupling with dual
boundary con-\linebreak dition.  In particular, it gives a dual
description of the two-dimensional degrees of freedom.

\section{The AGT correspondence}
\label{AGT}

The connection to the AGT correspondence is now clear.  The BRST
cohomology represented by the constrained affine currents is the
W-algebra $\CW_k(\gf)$, obtained from $\gfh$ of level $k$ by quantum
Drinfeld-Sokolov reduction \cite{Feigin:1990pn} with respect to the
principal embedding specified by the Nahm pole.  This is exactly the
symmetry that one finds on the two-dimensional side of the AGT
correspondence.

The level $k$ can be fixed by comparing the two effective descriptions
of the partition function of the $(2,0)$ theory, namely the Nekrasov
partition function of the $\CN = 2$ theory and the relevant conformal
block of the gauged WZW model.  This was the original idea of
\cite{Alday:2009aq}, and leads to the identification
\begin{equation}
  \label{k}
  k = -h^\vee - \frac{\epsilon_2}{\epsilon_1}.
\end{equation}
Notice that the six-dimensional description is symmetric under the
exchange of $\epsilon_1$ and $\epsilon_2$, which amounts to
$S$-duality in $\CN =4$ super Yang-Mills theory.  So the same must be
true for the resulting W-algebra.  There is indeed an isomorphism
\cite{Feigin:1991wy}
\begin{equation}
  \label{W-duality}
  \CW_k(\gf) \iso \CW_{k'}({}^L\gf)
\end{equation}
for any simple Lie algebra $\gf$, with the levels related by $k +
h^\vee = (k' + h^\vee)^{-1}$.  Here ${}^L\gf$ is the Langlands dual of
$\gf$, which in the simply-laced case is the same as $\gf$.

As an example, take $\gf = A_1$.  In this case the gauged WZW model is
Liouville theory and the W-algebra in question is the Virasoro
algebra.  More generally, for $\gf = A_{N-1}$, one gets a Toda theory
with symmetry algebra $\CW_N$, which contains the Virasoro algebra
with central charge
\begin{equation}
  c = N-1 + (N^3 - N) \frac{(\epsilon_1+\epsilon_2)^2}{\epsilon_1\epsilon_2}.
\end{equation}
The central charge exhibits the $N^3$ scaling behavior of the entropy
of the $(2,0)$ theory \cite{Klebanov:1996un}, reflecting the
six-dimensional origin of the two-dimensional degrees of freedom.

Let us see how the W-algebra structure is encoded in the physics of
the $\CN = 2$ theory.  To simplify the analysis, we content ourselves
with the semiclassical approximation which is good when $k$ is large.
Using the $G_-$ gauge symmetry one can put the affine currents
\linebreak into the form
\begin{equation}
  J_z/k = t_+ + \sum_{t_i \in \ker(t_-)} W_i t_i,
\end{equation}
where the sum is over the $t_a$ in the kernel of the adjoint action by
$t_- = t_1 - it_2$.  For $\gf = A_{N-1}$, the gauged-fixed form of $J$
is
\begin{equation}
  J_z/k = 
  \begin{pmatrix}
    0   & 1      & 0      & \dots & 0   & 0 \\
    0   & 0      & 1      & \dots & 0   & 0 \\
    \vdots & \vdots & \vdots & \ddots & \vdots & \vdots \\
    0   & 0      & 0      & \dots & 1   & 0 \\
    0   & 0      & 0      & \dots & 0   & 1 \\
    W_N & W_{N-1} & W_{N-2} & \dots & W_2 & 0
  \end{pmatrix},
\end{equation}
The classical W-algebra is generated by the currents $W_i$.

We consider the following polynomial in $x$:
\begin{equation}
  \label{lambda-J}
  \det(x - J_z/k).
\end{equation}
Its coefficients are elementary symmetric polynomials in the
eigenvalues of $J_z$, and as such can also be written as polynomials in
the Casimir operators $\Tr J_z^i$, $i = 2$, $\dots$, $\rank \gf$.  The
$\CN = 2$ theory compactified down to two dimensions using cigars is
equivalent to the $\CN = 4$ super Yang-Mills theory we have been
studying, compactified further on $C$.  The adjoint scalar $\phi$ from
the vector multiplet of the $\CN = 2$ theory is identified with $X_z$.
Noting that $J_z/k = X_z$ in the $Q$-cohomology since this holds at $x
= y = 0$, we see that the Casimirs are identified with the generators
$\Tr\phi^i$ of the chiral ring of the $\CN = 2$ theory.  The
coefficients of the polynomial \eqref{lambda-J} can of course be
expressed in terms of the $W_i$, so we obtain a relation between the
W-currents and these generators.  For example, for $\gf = A_{N-1}$ we
have
\begin{equation}
  \label{detlJ}
  \det(x - J_z/k)
  = x^N - \sum_{i=2}^N W_i x^{N-i},
\end{equation}
and we find the correspondence
\begin{equation}
  W_i \sim \Tr\phi^i + \dotsb.
\end{equation}
Essentially the same relation was proposed in \cite{Bonelli:2009zp}.

There is a more refined correspondence, relating the W-currents to the
Seiberg-Witten curve $\Sigma$ of the $\CN = 2$ theory which is a
branched cover of $C$.  If we choose coordinates $(x,z)$ for the
holomorphic cotangent bundle of $C$, then $\Sigma$ is given
\cite{Gaiotto:2009hg} in the absence of the $\Omega$-deformation by
\begin{equation}
  \vev{\det\bigl(x - X_z)  \dotsb}
  = \Bigl(x^N + \sum_{i=2}^N u_i(z) x^{N-i}\Bigr) \vev{\dotsb}
  = 0,
\end{equation}
where the ellipses denote the relevant defect operators inserted at
the punctures of $C$.  (The presence of the boundaries can be
neglected if $X_z$ is placed far away from them.)  The restriction of
the one-form $x dz$ to $\Sigma$ is the Seiberg-Witten differential.
Evaluating the correlation function at $x = y = 0$, we find
\begin{equation}
 \vev{W_i \dotsb} \sim u_i \vev{\dotsb}
\end{equation}
in the undeformed limit.  Hence, the expectation values of the
W-currents determine the Seiberg-Witten curve, from which the Coulomb
branch parameters can be read off.  This relation was conjectured in
\cite{Alday:2009aq}, and has been checked in \cite{Kanno:2009ga,
  Keller:2011ek}.

We have seen that our approach explains important aspects of the AGT
correspondence, that is, how W-algebras arise from the $(2,0)$ theory
in the $\Omega$-background and how they are related to the physics of
$\CN = 2$ gauge theories obtained by compactification on Riemann
surfaces.  Our construction can be generalized in a number of ways.

Clearly, it is desirable to treat all gauge groups in a uniform
fashion, not just simply-laced ones.  $\CN = 2$ theories with
non-simply-laced gauge groups can be constructed from the $(2,0)$
theory by compactification on Riemann surfaces with outer automorphism
twists \cite{Tachikawa:2010vg},\linebreak and there seems to be no
obstacle to adapting our construction to this situation.  The
sym-\linebreak metry algebras of the conformal field theories will
then be not ordinary W-algebras but twisted ones, and associated to
the Langlands dual of the affine Kac-Moody algebra of the gauge group
\cite{Braverman:2010ef, Keller:2011ek}.

Another possibility is to replace $\R^4$ by an ALE orbifold
$\C^2/\Z_k$.  In this case the symme-\linebreak try algebras are
parafermionic W-algebras for $k > 1$ \cite{Belavin:2011pp,
  Nishioka:2011jk}.  The main problem here will be to identify the
boundary conditions.  Once that is done, the same argument should lead
to a BRST construction of these algebras, generalizing quantum
Drinfeld-Sokolov reduction in the $k = 1$ case.

Finally, one may include a half-BPS surface operator in the $\CN = 2$
theory side \cite{Alday:2010vg, Kozcaz:2010yp, Braverman:2010ef,
  Wyllard:2010rp}.  This situation can be realized in our construction
by placing a codimension-two defect at the \linebreak origin of $D_2$.
The presence of the defect changes the residue of the Nahm pole to
another one \linebreak corresponding to a different $\suf(2)$
embedding \cite{Gaiotto:2008ak, Chacaltana:2012zy}, hence the
resulting W-algebra to the one associated to this new embedding.  So
everything we have said about the W-algebra sim-\linebreak ply carries
over to this case, except one important point: the duality of
W-algebras is lost, since the setup is no longer symmetric between
$D_1$ and $D_2$.  To remedy the asymmetry, we can place another defect
at the origin of $D_2$.  Then the setup is symmetric again, under the
\linebreak exchange of $\epsilon_1$ and $\epsilon_2$ together with the
exchange of the defects.  This consideration suggests that there is a
generalization of quantum Drinfeld-Sokolov reduction whose data are a
Lie algebra and two $\slf(2)$ embeddings, such that it enjoys an
analogous duality and reduces to the ordinary quantum Drinfeld-Sokolov
reduction when one of the embeddings is principal.

\subsection*{Acknowledgments}

I thank Nils Carqueville and Yuji Tachikawa for helpful comments.
This work is supported by Deutsche Forschungsgemeinschaft through the
Research Training Group 1670 ``Mathematics Inspired by String Theory
and QFT.''


\providecommand{\href}[2]{#2}\begingroup\raggedright\endgroup
\end{document}